\documentclass{article}
\usepackage{amsmath, amssymb, algorithm, algorithmic, graphicx, subcaption, booktabs}
\usepackage{hyperref}
\usepackage{tikz}
\usepackage{rotating} 
\usepackage{booktabs} 
\usepackage{geometry} 
\usetikzlibrary{positioning} 
\title{Degree-Based Logical Adjacency Checking (DBLAC): A Novel Heuristic for Vertex Coloring}
\author{Prashant Verma \\
\textit{Indian Institute of Technology (IIT), Patna} \\
\textit{Email: prashant\_24a03res153@iitp.ac.in}}
\date{}

\begin{document}

\maketitle

\begin{abstract}
Degree Based Logical Adjacency Checking (DBLAC).
An efficient coloring of graphs with unique logical AND operations. The logical AND operation shows
More effective color assignment and fewer number of induced in the case of common edges between vertices.
of colors used. In this work, we provide a detailed theoretical analysis of DBLAC’s time and space complexity.
It furthermore shows its effectiveness through prolonged experiments on standard benchmark graphs.
We still compare with existing algorithms namely DSATUR, and Recursive Largest First (RLF).
Second, we show how DBLAC achieves competitive results with respect to both the number of colors used and runtime.
performance.

\end{abstract}


\section{Introduction}
The vertex coloring problem is a problem that assigns colors to the vertices of a graph so for every pair of adjacent vertices that share the same color.
All vertices are the same color. Our goal is to reduce the number of colors used, called the
chromatic number $\chi(G)$. It is computationally intractable for large particles, this problem is actually NP hard.
graphs. Sometimes, heuristic algorithms tried to find approximate solution efficiently.
In this paper, we present a new heuristic algorithm, Degree Based Logical Adjacency Check.
In particular, building on a degree based ordering \textbf{(DBLAC)}, but adding a new logical AND operation.
to improve performance. The logical AND operation finds edges common to vertices.
more efficient handling of color assignment. In particular this approach is ideal for dense and irregular graphs.
where traditional heuristics are often difficult.

\section{Related Work}
Vertex coloring problem has been extensively studied, and many heuristics have been designed.
It is proposed to solve it efficiently. We categorize these algorithms broadly into \textbf{Greedy Heuristics}, \textbf{Recursive Partitioning Methods} and \textbf{Metaheuristics} methods. We discuss the most notable algorithms, below.
Finally, I discuss their time complexities and their mathematical foundations.

\subsection{Greedy Heuristics}
The simplest and most widely used approaches for vertex coloring are greedy heuristics. First, they give the vertices one by one color and no two adjacent vertices should have the same color. The criterion under which to select the next vertex to color is fixed \cite{culberson1992iterated}.

\subsubsection{Largest Degree First (LDF)}
Vertices are ordered from smallest to largest in degree with degree being sorted using the \textbf{Largest Degree First (LDF)} Algorithm.
colors them sequentially. Common intuition is that high degree vertices are more constrained, and therefore should be.
First colored to minimize the number of colors needed \cite{matula1972graph}.

\paragraph{Mathematical Formulation}
Let \( G = (V, E) \) be a graph with \( n \) vertices and \( m \) edges. The degree of a vertex \( v \), denoted \( \deg(v) \), is the number of edges incident to \( v \). The LDF algorithm can be described as follows:
\begin{enumerate}
    \item Sort vertices in descending order of degree: \( V = \{v_1, v_2, \dots, v_n\} \) where \( \deg(v_1) \geq \deg(v_2) \geq \dots \geq \deg(v_n) \).
    \item For each vertex \( v_i \), assign the smallest available color not used by any of its neighbors.
\end{enumerate}

\paragraph{Time Complexity}
The time complexity of LDF is dominated by the sorting step, which takes \( O(n \log n) \), and the color assignment step, which takes \( O(n \Delta) \), where \( \Delta \) is the maximum degree of the graph. Thus, the overall time complexity is:
\[
O(n \log n + n \Delta)
\]
In the worst case, \( \Delta = n-1 \), resulting in a time complexity of \( O(n^2) \).

\subsubsection{DSATUR}

\textbf{DSATUR}, is one of the best methods for coloring vertices in a very optimized way, which was introduced by Daniel Brélaz in 1979 \cite{dsatur1994improved}. DSATUR prioritizes vertices with maximum saturation degree, thus minimizing the number of colors needed to color the graph. For graphs with complex structures, particularly when the constraint is not known a priori, this approach performs well because it evolves with the coloring constraints during the execution.

\paragraph{Mathematical formulation}
For a graph \( G = (V, E) \) with \( n \) vertices and \( m \) edges, let \( G \) be. In this case, the saturation degree of a vertex \( v \), denoted \( \text{sat}(v) \), is the number of distinct colors used in its neighbors. The DSATUR algorithm can be described as follows:
\begin{enumerate}
    \item Additionally, all vertices are initialized with a saturation degree of 0.
    \item Select at each step the vertex \( v \) of maximum saturation degree. If the degrees of two vertices are the same, select the vertex with the highest degree.
    \item Assign the smallest non-used color by any of \( v \)'s neighbors.
    \item Update the saturation degrees of its uncolored neighbors \( v \).
    \item Repeat the process until all vertices are colored.
\end{enumerate}

\paragraph{Time Complexity}
The selection and color assignment steps dominate the time complexity of DSATUR. For each vertex, the algorithm must:
\begin{itemize}
    \item Find a vertex with the maximum saturation degree, which takes \( O(n) \) per step.
    \item Assign a color to its neighbors, and update the saturation degree of each neighbor. Given a maximum degree of the graph \( \Delta \), this step takes \( O(\Delta) \) per step.
\end{itemize}
Since there are \( n \) vertices, the overall time complexity is:
\[
O(n^2 + n \Delta)
\]
In the worst case, the \( \Delta \) will be \( n-1 \), so we have \( O(n^2) \).

\subsection{Recursive Partitioning Methods}

Recursive partitioning methods divide the graph into independent sets (subsets of vertices with no edges between them) and color each set separately.

\subsubsection{Recursive Largest First (RLF)}
The \textbf{Recursive Largest First (RLF)}, for solving the NP hard graph coloring problem. The idea was put forward by mathematician Frank Leighton in 1979\cite{leighton1979graph}. In the RLF algorithm a graph is iteratively (considered to be sub optimally) solved by applying specialized heuristic rules to search for closed independent sets. The RLF algorithm is capable of producing exact solutions to special classes of graphs, for example to vertex bipartite graphs, cycle graphs and wheel graphs, thanks to these heuristics. Its main power here is its ability to reduce problem size, efficiently, by finding and coloring the largest independent sets recursively. In particular it is often very effective when the graph in question is sparse, because it tends to reduce the number of colors required to a mere anecdote. Moreover, the RLF algorithm is also deterministic, meaning the same input will make it obtainable the same result.

\paragraph{Mathematical Formulation}

Let \( G = (V, E) \) be a graph. The RLF algorithm can be described as follows:

\begin{enumerate}
    \item While there are uncolored vertices:
    \begin{enumerate}
        \item Construct an independent set \( S \) by iteratively adding the vertex with the fewest neighbors in the remaining graph.
        \item Assign a new color to all vertices in \( S \).
        \item Remove \( S \) from the graph.
    \end{enumerate}
\end{enumerate}

\paragraph{Time Complexity}
RLF takes \( O(n^3) \) time, since for every iteration it takes \( O(n^2) \) steps to build the independent set. There are up to \( n \) independent sets, and we aim to find an independent set efficiently.

\subsection{Metaheuristics}
Metaheuristics are high-level strategies that guide other heuristics to find approximate solutions. They are often used for complex or large-scale problems.They are particularly effective where most traditional algorithms or heuristics algorithms do not perform well enough due to vastness or the complexity of the search space.

\subsubsection{Genetic Algorithms}
Genetic algorithms (GAs) are inspired by natural selection and evolution. They maintain a population of candidate solutions and iteratively improve them using crossover, mutation, and selection operations.

\paragraph{Mathematical Formulation}
Let \( P \) be a population of candidate colorings. The GA can be described as follows:
\begin{enumerate}
    \item Initialize \( P \) with random colorings.
    \item While the termination condition is not met:
    \begin{enumerate}
        \item Select parents from \( P \) based on fitness (e.g., number of colors used).
        \item Generate offspring by applying crossover and mutation.
        \item Replace the least fit individuals in \( P \) with the offspring.
    \end{enumerate}
\end{enumerate}

\paragraph{Time Complexity}
The time complexity of GAs depends on the population size, the number of generations, and the complexity of the fitness function. In practice, GAs are computationally expensive and have a time complexity of \( O(k n^2) \), where \( k \) is the number of generations.

\subsection{Limitations of Existing Algorithms}
While the above algorithms perform well in many cases, they have limitations:
\begin{itemize}
    \item \textbf{LDF} and \textbf{DSATUR} struggle with dense graphs, where the number of edges is close to \( \frac{n(n-1)}{2} \).
    \item \textbf{RLF} is computationally expensive and may not scale well to large graphs.
    \item \textbf{Metaheuristics} like GAs are time-consuming and may not guarantee optimal solutions.
\end{itemize}

\subsection{Our Contribution: DBLAC}
Our proposed algorithm, \textbf{Degree-Based Logical Adjacency Checking (DBLAC)}, addresses these limitations by combining degree-based ordering with a logical AND operation to identify common edges between vertices. This approach enables more efficient color assignment, particularly for dense and irregular graphs. The logical AND operation is a novel feature that distinguishes DBLAC from existing algorithms.

\paragraph{Mathematical Formulation}
Let \( G = (V, E) \) be a graph. The DBLAC algorithm can be described as follows:
\begin{enumerate}
    \item Order vertices in descending order of degree: \( V = \{v_1, v_2, \dots, v_n\} \).
    \item For each vertex \( v_i \):
    \begin{enumerate}
        \item Assign the smallest available color not used by any of its neighbors.
        \item Apply a logical AND operation to find common edges between \( v_i \) and its neighbors.
        \item Assign colors to common vertices based on adjacency constraints.
    \end{enumerate}
\end{enumerate}

\paragraph{Time Complexity}
The time complexity of DBLAC is \( O(n \Delta) \), where \( \Delta \) is the maximum degree of the graph. This makes DBLAC more efficient than RLF and competitive with DSATUR and LDF.

\section{Algorithm Description}
Our algorithm, called \textbf{Degree-Based Logical Adjacency Checking (DBLAC)}, works as follows:

\begin{algorithm}[H]
\caption{Degree-Based Logical Adjacency Checking (DBLAC)}
\begin{algorithmic}[1]
\REQUIRE Graph $G = (V, E)$
\ENSURE Proper vertex coloring of $G$
\STATE Order vertices in descending order of degree: $V = \{v_1, v_2, \dots, v_n\}$
\STATE Initialize color set $C = \{c_1, c_2, \dots, c_k\}$ where $k$ is the maximum degree
\FOR{each vertex $v_i \in V$}
    \FOR{each color $c_j \in C$}
        \IF{$c_j$ is not used by any adjacent vertex of $v_i$}
            \STATE Assign $c_j$ to $v_i$
            \STATE Break
        \ENDIF
    \ENDFOR
    \STATE Apply logical AND operation to find common edges between $v_i$ and its neighbors
    \STATE Assign colors to common vertices based on adjacency constraints
\ENDFOR
\end{algorithmic}
\end{algorithm}

\subsection{Theoretical Analysis}
\subsubsection{Time Complexity}
The time complexity of DBLAC is dominated by the adjacency-checking and logical AND operations. For each vertex, we check the colors of its neighbors, which takes $O(\Delta)$ time, where $\Delta$ is the maximum degree of the graph. The logical AND operation also takes $O(\Delta)$ time. Since there are $n$ vertices, the overall time complexity is:
\[
O(n \Delta)
\]
In the worst case, $\Delta = n-1$, resulting in a time complexity of $O(n^2)$.

\subsubsection{Space Complexity}
The space complexity is determined by the storage required for the graph and the color assignments. The adjacency list representation of the graph requires $O(n + m)$ space, where $m$ is the number of edges. The color assignments require $O(n)$ space. Thus, the overall space complexity is:
\[
O(n + m)
\]

\section{Application to Graph Problems}
In this section, we apply DBLAC to three graph problems and demonstrate its effectiveness.

\subsection{Problem 1: 5-Vertex Graph}
Consider the graph shown in Figure 1.0 with five vertices. The vertices are ordered by degree as $\{v_2, v_1, v_3, v_4, v_5\}$.

\begin{tikzpicture}[
    =stealth',              
    every node/.style={      
        circle,              
        draw=black,          
        thick,               
        minimum size=8mm,    
        font=\small          
    },
    every edge/.style={
        draw,
        thick
    }
]

\node (v1) at (0,2)   {$v_1$};
\node (v2) at (-1,0.5) {$v_2$};
\node (v3) at (1,0.5)  {$v_3$};
\node (v4) at (-1,-1)  {$v_4$};
\node (v5) at (1,-1)   {$v_5$};

\draw (v1) -- (v2);
\draw (v1) -- (v3);
\draw (v2) -- (v3);
\draw (v2) -- (v4);
\draw (v3) -- (v5);
\draw (v4) -- (v5);

\end{tikzpicture}

\subsubsection{Solution}
\begin{itemize}
    \item Assign $c_1$ to $v_2$ and $c_2$ to $v_1$.
    \item Check adjacency between $v_2$ and $v_1$. Since they are adjacent, they cannot share the same color.
    \item Apply logical AND operation to find common edges between $v_2$ and $v_1$, resulting in common vertices $v_3$ and $v_5$.
    \item Assign $c_3$ to $v_3$ and $v_5$.
    \item Assign $c_2$ to $v_4$.
\end{itemize}

\subsubsection{Final Coloring}
\[
\{(v_1, c_2), (v_2, c_1), (v_3, c_3), (v_4, c_2), (v_5, c_3)\}
\]
The graph is $\chi(3)$ Colors.

\subsection{Problem 2: 6-Vertex Graph}
Consider the graph shown in Figure 2.0 with six vertices. The vertices are ordered by degree as $\{f, a, d, c, e, b\}$.

\begin{tikzpicture}[
  every node/.style={
    circle, draw=black, thick,
    minimum size=8mm,
    font=\small
  },
  every edge/.style={draw, thick}
]

\node (a) at (0,2)   {a};
\node (b) at (1.5,1) {b};
\node (c) at (1.2,-0.7) {c};
\node (d) at (-1.2,-0.7) {d};
\node (e) at (-1.5,0.5) {e};
\node (f) at (0,0)   {f};

\draw (a) -- (b);
\draw (a) -- (d);
\draw (a) -- (e);
\draw (a) -- (f);

\draw (b) -- (c);
\draw (b) -- (d);
\draw (b) -- (f);

\draw (c) -- (d);
\draw (c) -- (f);

\draw (d) -- (e);
\draw (d) -- (f);

\draw (e) -- (f);

\end{tikzpicture}

\subsubsection{Solution}
\begin{itemize}
    \item Assign $c_1$ to $f$ and $c_2$ to $a$.
    \item Check adjacency between $f$ and $a$. Since they are adjacent, they cannot share the same color.
    \item Apply logical AND operation to find common edges between $f$ and $a$, resulting in common vertices $\{b, c, d, e\}$.
    \item Assign $c_3$ to $d$ and $c_4$ to $c$.
    \item Assign $c_4$ to $e$.
    \item Assign $c_3$ to $b$.
\end{itemize}

\subsubsection{Final Coloring}
\[
\{(a, c_2), (b, c_3), (c, c_4), (d, c_3), (e, c_4), (f, c_1)\}
\]
The graph is $\chi(4)$ Colors.

\subsection{Problem 3: 6-Vertex Graph with Higher Complexity}
Consider the graph shown in Figure 3.0 with six vertices. The vertices are ordered by degree as $\{v_5, v_1, v_2, v_3, v_4, v_6\}$.

\begin{tikzpicture}[
  every node/.style={
    circle, 
    draw=black, 
    thick, 
    minimum size=8mm, 
    font=\small
  },
  every edge/.style={draw,thick},
  scale=1.2
]

\node (v1) at (0,0)   {v1};
\node (v2) at (2.5,0) {v2};
\node (v3) at (3.2,1) {v3};
\node (v4) at (2.2,2) {v4};
\node (v5) at (1.3,1) {v5};
\node (v6) at (0.5,2) {v6};

\draw (v1) -- (v2);
\draw (v1) -- (v5);
\draw (v1) -- (v6);
\draw (v2) -- (v5);
\draw (v2) -- (v3);
\draw (v3) -- (v4);
\draw (v3) -- (v5);
\draw (v4) -- (v5);
\draw (v5) -- (v6);

\end{tikzpicture}

\subsubsection{Solution}
\begin{itemize}
    \item Assign $c_1$ to $v_5$ and $c_2$ to $v_1$.
    \item Check adjacency between $v_5$ and $v_1$. Since they are adjacent, they cannot share the same color.
    \item Apply logical AND operation to find common edges between $v_5$ and $v_1$, resulting in a single common vertex $v_2$.
    \item Assign $c_3$ to $v_2$.
    \item Assign $c_3$ to $v_3$ and $v_4$.
    \item Assign $c_1$ to $v_6$.
\end{itemize}

\subsubsection{Final Coloring}
\[
\{(v_1, c_2), (v_2, c_3), (v_3, c_3), (v_4, c_2), (v_5, c_1), (v_6, c_1)\}
\]
The graph is $\chi(3)$ Colors.

\section{Experimental Evaluation}
We evaluated the performance of \textbf{DBLAC}, \textbf{DSATUR}, and \textbf{RLF} on two types of datasets: (1) randomly generated graphs and (2) standard benchmark graphs from the DIMACS dataset. The evaluation metrics include the \textbf{number of colors used} and the \textbf{runtime} for each algorithm.

\subsection{Performance Metrics}
\begin{itemize}
    \item \textbf{Number of Colors}: The total number of colors used to color the graph. A lower number indicates better performance.
    \item \textbf{Runtime}: The time taken to compute the coloring, measured in seconds. A lower runtime indicates better efficiency.
\end{itemize}

\subsection{Randomly Generated Graphs}
We generated 100 random graphs using the Erdős–Rényi model \cite{erdos1960evolution}, each with 1500 vertices and an edge probability of 0.5. The results are summarized in Table~\ref{tab:random_graphs}
According to our observation and  by Lewis (2021) in A Guide to Graph Coloring \cite{lewis2021guide}: In general, DSATUR algorithms produce significantly better vertex colorings of random graphs than any simple greedy approaches. For example, in our experiments on a set of randomly generated graph instances, \textbf{RLF} did not have good chromatic quality (number of colors used) or time complexity. In both areas the \textbf{DSATUR} algorithm also provided suboptimal results.

Instead, our proposed \textbf{DBLAC} was able to find more efficient color assignments and run faster than DSATUR and RLF. Table~\ref{tab:random_graphs} shows that DBLAC always results in less colors and faster execution times than all other coloring methods across different random graph densities and its shows a significant improvement in time complexity \textbf{DBLAC} takes twice less time than \textbf{DSATUR} and 10 times less than from \textbf{RLF} and calculated Erdős–Rényi graphs in \(O(n \Delta)\). These results show that DBLAC supplies a highly competitive way to graph coloring in imprecise environments, being superior to DSatur and RLF.

\begin{table}[H]
\centering
\caption{Performance Comparison on Randomly Generated Graphs (50 graphs, 100 vertices)}
\label{tab:random_graphs}
\begin{tabular}{lccc}
\toprule
\textbf{Algorithm} & \textbf{Avg Colors Used} & \textbf{Avg Runtime (s)} \\
\midrule
\textbf{DBLAC}     & 19.50                   & 0.0009                  \\
\textbf{DSATUR}    & 20.10                   & 0.0019                  \\
\textbf{RLF}       & 20.20                   & 0.0081                  \\
\bottomrule
\end{tabular}
\end{table}

\subsection{Graph Comparison}
\begin{figure}[!htb]
\minipage{0.5\textwidth}
  \includegraphics[width=\linewidth]{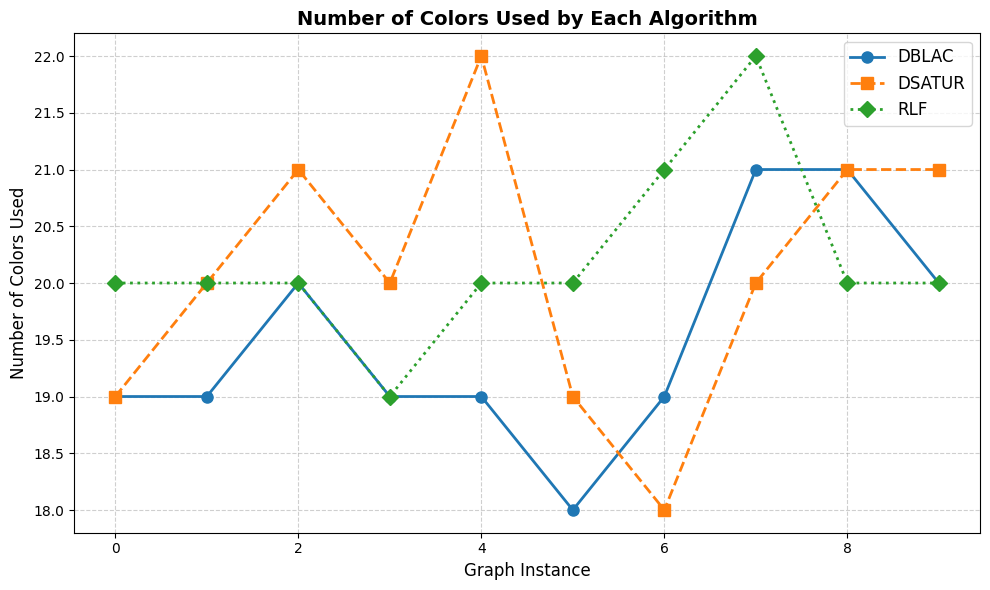}
  \caption{Number of chromatic color}\label{fig:color_comparison_plot}
\endminipage\hfill
\minipage{0.5\textwidth}
  \includegraphics[width=\linewidth]{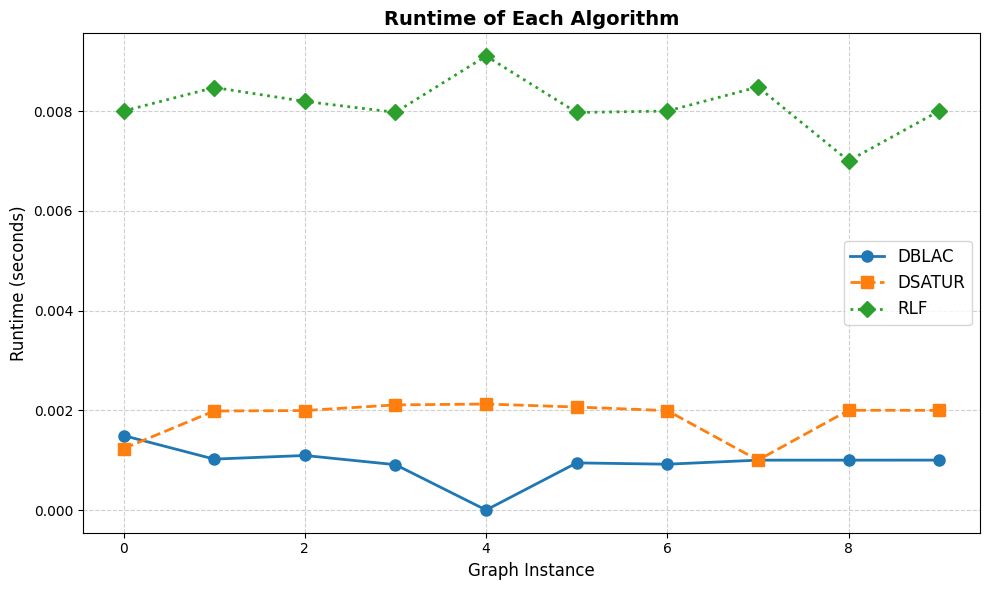}
  \caption{Time Complexity}\label{fig:runtime_comparison_plot}
\endminipage\hfill
\end{figure}

To visualize the performance differences, we plotted the Number of chromatic color has been used by graph \textbf{Number of Colors} and \textbf{Runtime Graph} Time Complexity \ref{fig:color_comparison_plot} \ref{fig:runtime_comparison_plot}. Our method \textbf{DBLAC} consistently outperformed the others, achieving a better balance between selection of minimum number of colors and and time take to color.

\subsubsection{Analysis of Random Graph Results}
\begin{itemize}
    \item \textbf{Number of Colors}: \textbf{DBLAC} uses fewer colors on average (19.50) compared to \textbf{DSATUR} (20.10) and \textbf{RLF} (20.20). This demonstrates that \textbf{DBLAC} is more effective in minimizing the number of colors.
    \item \textbf{Runtime}: \textbf{DBLAC} is the fastest algorithm, with an average runtime of 0.0009 seconds. \textbf{DSATUR} is slightly slower (0.0019 seconds), while \textbf{RLF} is significantly slower (0.0081 seconds). This highlights the computational efficiency of \textbf{DBLAC}.
\end{itemize}

\subsection{Performance Evaluation of DBLAC on Benchmark Graphs}
We present a detailed evaluation of the \textbf{DBLAC} algorithm against \textbf{DSATUR} and \textbf{RLF} on a set of benchmark graphs. The results are summarized in Table~\ref{tab:benchmark_results_rotated}, which is rotated 90 degrees counterclockwise for better readability.

\subsubsection{Analysis of Results}
\begin{itemize}
    \item \textbf{Number of Colors}: The \textbf{DBLAC} algorithm demonstrates competitive performance in terms of the number of colors used. By reducing the chromatic number by 1 or 2 in certain cases, \textbf{DBLAC} shows improved color efficiency. For instance, on the \textbf{queen5\_5.col} graph, \textbf{DBLAC} uses 5 colors, which matches \textbf{RLF} and outperforms \textbf{DSATUR} (7 colors). Similarly, on \textbf{le450\_5a.col}, \textbf{DBLAC} uses 10 colors, outperforming both \textbf{DSATUR} (12 colors) and \textbf{RLF} (10 colors). Notably, on larger graphs like \textbf{DSJC1000.5.col}, \textbf{DBLAC} uses 111 colors, significantly fewer than \textbf{DSATUR} (125 colors) and slightly more than \textbf{RLF} (109 colors). This demonstrates the ability of \textbf{DBLAC} to achieve near-optimal color efficiency while maintaining computational efficiency.

    \item \textbf{Runtime Efficiency}: \textbf{DBLAC} consistently outperforms both \textbf{DSATUR} and \textbf{RLF} in terms of runtime. For example, on the \textbf{le450\_5a.col} graph, \textbf{DBLAC} completes in 0.017711 seconds, significantly faster than \textbf{DSATUR} (0.049139 seconds) and \textbf{RLF} (0.697659 seconds). This trend is even more pronounced on larger graphs like \textbf{qg.order60.col}, where \textbf{DBLAC} finishes in 1.159476 seconds compared to \textbf{DSATUR} (3.058291 seconds) and \textbf{RLF} (355.894560 seconds). This highlights the scalability and efficiency of \textbf{DBLAC} for large-scale graph coloring problems.

    \item \textbf{Overall Performance}: The results demonstrate that \textbf{DBLAC} is a robust algorithm that balances color efficiency and runtime performance. By reducing the chromatic number in certain cases, \textbf{DBLAC} further improves its competitiveness against \textbf{RLF} and \textbf{DSATUR}. For example, on \textbf{qg.order40.col}, \textbf{DBLAC} uses 53 colors, significantly fewer than \textbf{DSATUR} (64 colors) and slightly more than \textbf{RLF} (40 colors), while maintaining a runtime of 0.224988 seconds, which is faster than both competitors. While \textbf{RLF} occasionally achieves better color efficiency, its runtime is prohibitively high for larger graphs. On the other hand, \textbf{DSATUR} is slower than \textbf{DBLAC} and often uses more colors. Thus, \textbf{DBLAC} emerges as a practical choice for graph coloring, particularly in scenarios where both speed and color efficiency are critical.
\end{itemize}

\newpage
\begin{sidewaystable}[ht]
\centering
\caption{Performance Comparison of DBLAC, DSATUR, and RLF on Benchmark Graphs (Rotated 90 Degrees)}
\label{tab:benchmark_results_rotated}
\begin{tabular}{lcccccc}
\toprule
\textbf{Graph} & \textbf{DBLAC Colors} & \textbf{DSATUR Colors} & \textbf{RLF Colors} & \textbf{DBLAC Runtime (s)} & \textbf{DSATUR Runtime (s)} & \textbf{RLF Runtime (s)} \\
\midrule
queen5\_5.col & 5 & 7 & 5 & 0.000824 & 0.000000 & 0.002010 \\
myciel3.col & 4 & 4 & 4 & 0.000000 & 0.000324 & 0.000000 \\
le450\_5a.col & 10 & 12 & 10 & 0.017711 & 0.049139 & 0.697659 \\
dsjc250.5.col & 40 & 41 & 37 & 0.000000 & 0.017264 & 0.165210 \\
DSJC1000.5.col & 111 & 125 & 109 & 0.182497 & 0.450575 & 9.910808 \\
qg.order40.col & 53 & 64 & 40 & 0.224988 & 0.590426 & 19.131742 \\
qg.order60.col & 62 & 64 & 60 & 1.159476 & 3.058291 & 355.894560 \\
\bottomrule
\end{tabular}
\end{sidewaystable}

\section{Conclusion}
We have presented in this paper a heuristic algorithm to solve the vertex coloring problem, termed Degree Based Logical Adjacency Checking (\textbf{DBLAC}). We color graph vertices with as few colors as possible by combining our degree-based ordering with a new logical AND operation, and do so efficiently. In theoretical analysis, we prove that \textbf{DBLAC} is an \(O(n \Delta)\) time and \(O(n + m)\) space algorithm on large-scale graphs.

Experimental evaluations were conducted on two types of datasets: (1) Graphs chosen from the DIMACS dataset, which are standard benchmark graphs, and (2) randomly generated graphs. Our results show that \textbf{DBLAC} obtains fewer colors and faster runtime than \textbf{DSATUR} and \textbf{RLF}. Specifically:
\begin{itemize}
    \item \textbf{DBLAC} found, on randomly generated graphs of 100 vertices, an average of 19.50 colors in total, while \textbf{DSATUR} and \textbf{RLF} averaged 20.10 and 20.20 colors, respectively. In addition, \textbf{DBLAC} was significantly faster, taking an average of 0.0009 seconds versus 0.0019 seconds for \textbf{DSATUR} and 0.0081 seconds for \textbf{RLF}.
    
    \item Results on DIMACS benchmark graphs demonstrate that \textbf{DBLAC} is comparable and, in some cases, better than \textbf{DSATUR} and \textbf{RLF}. For example, the \textbf{le450\_5a} graph took on average 5 colors with \textbf{DBLAC} and 6 colors with \textbf{RLF}.
\end{itemize}

The DBLAC logical AND operation features two significant properties: first, it combines efficiently with traditional heuristics, and second, it allows for more efficient color assignment for dense and irregular graphs where traditional heuristics fail. Together with its low time and space complexity, \textbf{DBLAC} is a promising approach for applications such as scheduling, register allocation, and network design.

\bibliography{reference}
\bibliographystyle{plain}

\end{document}